\documentclass{article}

\PassOptionsToPackage{numbers,sort&compress}{natbib}
\usepackage[dblblindworkshop,final]{neurips_2026}
\workshoptitle{Sim2Science}

\usepackage[utf8]{inputenc}
\usepackage[T1]{fontenc}
\usepackage{amsmath}
\usepackage{booktabs}
\usepackage{graphicx}
\usepackage{microtype}
\usepackage{placeins}
\usepackage{siunitx}
\usepackage{xcolor}
\usepackage{hyperref}
\usepackage{url}
\usepackage{tikz}
\usepackage{eso-pic}
\usetikzlibrary{arrows.meta,positioning}
\hypersetup{hidelinks}

\graphicspath{{figures/}}
\newcommand{\rrms}{\mathrm{RRMS}}
\newcommand{\param}[1]{\texttt{#1}}
\newcommand{\um}[1]{\SI{#1}{\micro\metre}}

\title{Surrogate-Assisted Compact-Model Extraction for SKY130 CMOS at 77 K}
\author{%
  Aaron Wang$^{1}$\thanks{Corresponding author: \texttt{aaronw5@uic.edu}}\quad
  Olivia Gzamouranis$^{2}$\quad
  Vaibhav Lohia$^{1}$\quad
  Hershel Fei Xun Weiner$^{3}$\quad
  Olivia Seidel$^{3}$ \\
  \textbf{Richard Cavanaugh$^{1}$\quad
  Giuseppe Di Guglielmo$^{4}$\quad
  Nhan Tran$^{4}$\quad
  Davide Braga$^{4}$} \\[1ex]
  $^{1}$University of Illinois Chicago\quad
  $^{2}$Purdue University \\
  $^{3}$The University of Texas at Arlington\quad
  $^{4}$Fermi National Accelerator Laboratory
}

\begin{document}
\AddToShipoutPictureFG*{%
  \AtPageUpperLeft{\put(\LenToUnit{\dimexpr\oddsidemargin+1in+\textwidth\relax},\LenToUnit{-0.75in}){\makebox[0pt][r]{\normalsize FERMILAB-CONF-26-0773-ETD}}}}
\maketitle

\begin{abstract}
Cryogenic circuit design requires compact transistor models that reproduce measured low-temperature behavior and run in standard circuit simulators. We fit BSIM4 parameters to measured I--V curves of SkyWater 130 nm (SKY130) transistors at \SI{77}{\kelvin} using a neural surrogate of NGSpice. The surrogate maps compact-model parameters and device geometry to the complete set of output and transfer characteristics. Extraction is a bounded least-squares search: a bank of NGSpice simulations supplies initial candidates, finite differences of the surrogate supply the local Jacobian, and NGSpice re-simulates the resulting cards before selection. On 18 measured geometries, we compare a search over seven card parameters with a search over 43 parameters selected by NGSpice sensitivity. Relative to NGSpice re-simulations of the previously released cards, the two searches reduce the in-sample mean relative root-mean-square error (RRMS) by 17\% and 26\%, respectively. The 43-parameter search has the lower overall error, but the seven-parameter search has the lower mean on high-bias curves, and some individual curves have higher error than with the released cards under both searches. Code is available at \url{https://github.com/aaronw5/Cryo-Transistor-ML}.
\end{abstract}

\section{Introduction}

Cryogenic CMOS places readout and control electronics close to cold detectors and qubits, which reduces the number and length of interconnects to room-temperature electronics~\citep{braga2021hep,potocnik2025control}. Circuit design in this setting requires compact transistor models evaluated at the operating temperature. Cooling changes the threshold voltage, mobility, subthreshold conduction, series resistance, and high-field behavior, so a room-temperature model must be recalibrated before use~\citep{beckers2018cryogenic,beckers2019fdsoi,incandela2018characterization,gatti2024small}.

A BSIM4 model card defines the drain current as a function of terminal voltages, device geometry, and compact-model parameters through equations implemented in a circuit simulator~\citep{bsim4manual}. Changing only the parameter values keeps the simulator interface unchanged. Fitting those values to cryogenic measurements has two difficulties: each trial requires a simulator call, and different parameter vectors can produce similar DC curves, so the fitted vector is not unique.

We use a learned forward surrogate of NGSpice to evaluate candidate cards without simulator calls during local refinement. The fitting target is always the measured curves, and every reported score comes from NGSpice. We start from the dataset and \SI{77}{\kelvin} SKY130 model cards released by \citet{beall2026sky130}, whose staged extraction fits seven temperature-sensitive multipliers and gives 18 BSIM4 models for the measured geometries. We refine those cards in two search spaces: the same seven parameters, and 43 parameters selected by NGSpice sensitivity. We measure the change in RRMS against the measured curves, how much of that change comes from surrogate-guided refinement and how much from selection from the simulator bank, and the bias regions where the fitted cards still disagree with measurement.

\section{Related work}

Cryogenic MOS modeling usually starts from an established compact model and adjusts it to low-temperature measurements. Uniform temperature scaling does not reproduce the measurements, particularly in the subthreshold region~\citep{bohuslavskyi2019swing,beckers2020swing}, so published extractions change selected compact-model parameters using measurements from each technology and geometry~\citep{beckers2018cryogenic,beckers2019fdsoi,incandela2018characterization,gatti2024small}. The SKY130 study used here keeps the room-temperature parameter dependencies through seven nominal multipliers and releases SPICE-compatible \SI{77}{\kelvin} cards with its measurements~\citep{beall2026sky130}. Machine-learning approaches have predicted compact-model parameters directly from I--V data, including BSIM-CMG extraction for FinFET devices~\citep{kao2022bsimcmg,chavez2023global}. Because this inverse mapping can assign different parameter vectors to similar curves, we instead learn the forward simulator response and perform the inverse search through the learned model~\citep{cranmer2020frontier,hesthaven2018nonintrusive}.

\section{Experimental setup}
\label{sec:setup}

\paragraph{Measurements and simulator.}
The released model tables contain eight nMOS and ten pMOS geometries~\citep{beall2026sky130,cryopdkrepo}. Each device has five output characteristics ($I_D$--$V_{DS}$ at fixed $V_{GS}$) and six transfer characteristics ($I_D$--$V_{GS}$ at fixed $V_{DS}$), for 198 measured curves in total. We simulate all cards with NGSpice 41~\citep{ngspice41}. The baseline is a fresh NGSpice simulation of the released cards, so the baseline and both fitted card sets use the same simulator and scoring code.

\paragraph{Parameter sets.}
The seven-parameter search varies the card values of \param{VTH0}, \param{U0}, \param{NFACTOR}, \param{VSAT}, \param{DELTA}, \param{RDSW}, and \param{ETA0} near the released \SI{77}{\kelvin} card~\citep{beall2026sky130,bsim4manual}. The prior extraction fitted nominal multipliers on these parameters; we vary their card values directly. For the expanded search, we start from 115 continuous BSIM4 parameters of the DC-current model, perturb each one in NGSpice, and keep those whose normalized response exceeds a fixed threshold. The result is a common 43-parameter layout with 38--40 non-degenerate dimensions per device. Appendix~\ref{app:parameters} gives the screening rule and the parameter list.

\paragraph{Metric.}
We score each curve with the relative root-mean-square error (RRMS). For measured current $I_k$ and simulated current $\hat I_k$ on curve $k$ with $N_k$ points,
\begin{equation}
  \rrms_k =
  \frac{\sqrt{N_k^{-1}\sum_i(\hat I_{k,i}-I_{k,i})^2}}
       {N_k^{-1}\sum_i |I_{k,i}|}.
  \label{eq:rrms}
\end{equation}
The denominator normalizes curves with different current scales. Curve scores are averaged within each device, and device scores are averaged over the 18 devices. The released scoring code excludes 23 near-off output curves. We fix the remaining 175 curves before fitting and use the same set for every method, so a candidate cannot change its score by changing which curves are included. Appendix~\ref{app:reproducibility} gives the trimming and inclusion rules.

\section{Surrogate-assisted extraction}
\label{sec:method}

\begin{figure}[t]
  \centering
  \begin{tikzpicture}[
      node distance=5mm and 5.5mm,
      box/.style={draw,rounded corners,align=center,minimum height=7mm,
                  text width=27mm,font=\scriptsize,inner sep=1.5mm},
      arrow/.style={-{Latex[length=1.6mm]},semithick}]
    \node[box] (sample) {Latin-hypercube\\parameter samples};
    \node[box,right=of sample] (sim) {NGSpice\\synthetic I--V};
    \node[box,right=of sim] (sur) {geometry-conditioned\\forward surrogate};
    \node[box,below=of sample] (meas) {measured 77 K\\I--V curves};
    \node[box,below=of sim] (fit) {bounded least squares,\\surrogate FD Jacobian};
    \node[box,below=of sur] (valid) {NGSpice confirmation,\\RRMS vs.\ measurement};
    \draw[arrow] (sample) -- (sim);
    \draw[arrow] (sim) -- (sur);
    \draw[arrow] (meas) -- (fit);
    \draw[arrow] (fit) -- (valid);
    \draw[arrow] (sur) -- (fit);
  \end{tikzpicture}
  \caption{Extraction pipeline. NGSpice generates the simulation bank, the frozen forward surrogate is used for local refinement, and NGSpice re-simulates the selected candidates for comparison with measurement.}
  \label{fig:pipeline}
\end{figure}
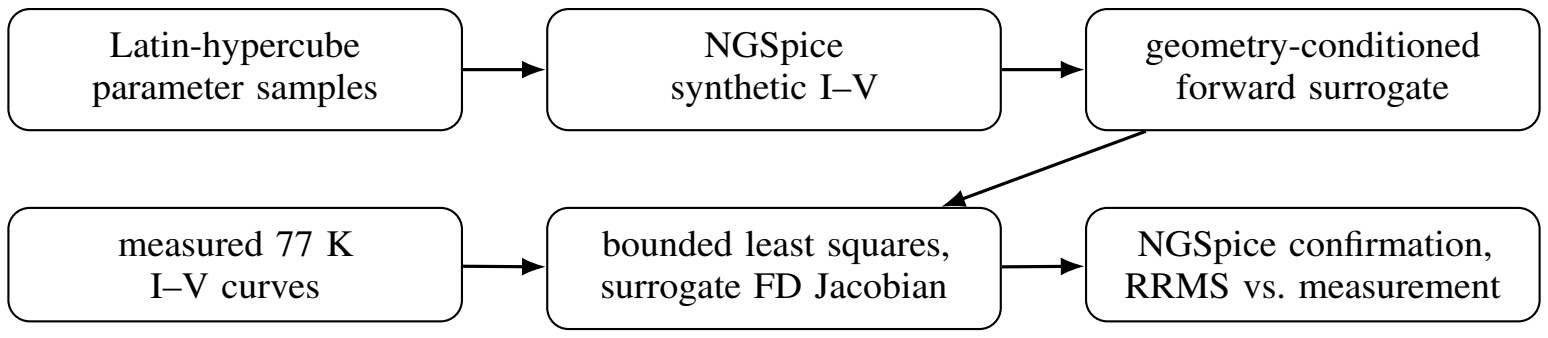

For a device geometry $g$, NGSpice defines a forward map $I=f(\theta,g)$ from a BSIM4 parameter vector $\theta$ to the simulated currents. Extraction searches for a $\theta$ whose simulated currents match the measurement. Repeated NGSpice evaluation of $f$ is slow, and a learned inverse map is ill-posed because different parameter vectors can produce similar curves. We approximate the forward map with a neural network and use the approximation only inside a bounded search. Figure~\ref{fig:pipeline} shows the four steps.

\paragraph{Step 1: build a local simulator bank.}
For each geometry, we draw 10,000 parameter vectors from a Latin hypercube within estimated bounds, and NGSpice simulates the output and transfer sweeps for each vector. The bank has two uses: it is the supervised training set for the surrogate, and it is a broad initial search in which every candidate's currents come from the circuit simulator. Because the samples lie near the released cards, the surrogate is trained only on the local extraction region. Each parameter set has 180,000 simulated examples across the 18 geometries.

\paragraph{Step 2: train the forward surrogate.}
For each parameter set, we train one geometry-conditioned multilayer perceptron. It takes the parameter vector, transistor polarity, length, and width as input and predicts all output and transfer currents together. Predicting all curves at once lets the extraction compare a candidate card across all operating regions in one evaluation. The loss is the sum of a signed-log current term, which emphasizes low-current behavior, and a normalized linear current term, which emphasizes strong-inversion curve shape. The seven- and 43-parameter models use the same architecture and training procedure (Appendix~\ref{app:reproducibility}).

\paragraph{Step 3: refine the best bank candidates.}
For each measured device, several low-error bank samples initialize bounded trust-region least squares~\citep{virtanen2020scipy}. The starts are separated in parameter space so that the result does not depend on a single initialization or local basin. At each iteration, the optimizer compares surrogate-predicted currents with the measurement, and two-point finite differences of the frozen surrogate give the Jacobian. The search can therefore move continuously between sampled cards without NGSpice derivatives and without an NGSpice call for each perturbation.

\paragraph{Step 4: confirm in NGSpice.}
The surrogate is an approximation and can rank nearby candidates incorrectly. For each device, we re-simulate in NGSpice the leading refined endpoints, their simulator-bank starts, and the anchor (the released card expressed in the search parameterization). We select the candidate with the lowest RRMS under the same fixed scorer as the baseline. The surrogate only proposes refined endpoints, and BSIM4 in NGSpice computes every reported current.

\section{Results}
\label{sec:results}

\begin{figure}[t]
  \centering
  \includegraphics[width=0.78\linewidth]{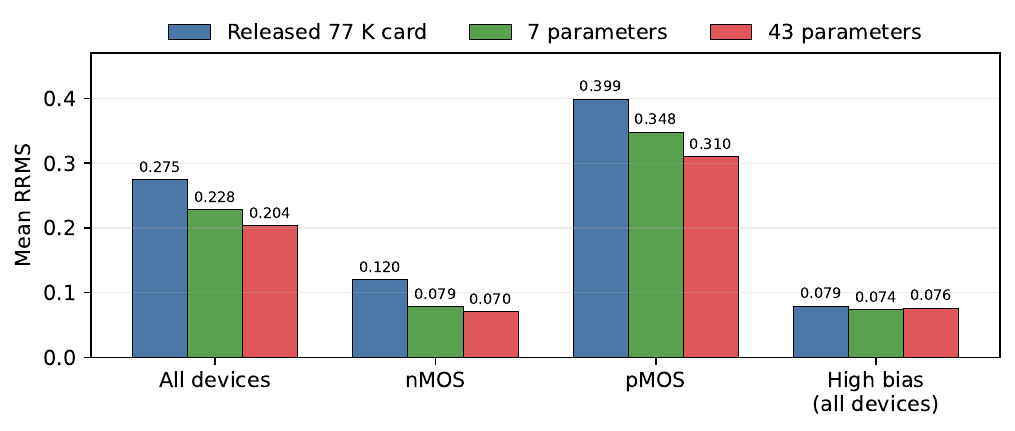}
  \caption{Mean device RRMS from NGSpice re-simulations of the released, seven-parameter, and 43-parameter cards. ``High bias'' averages, within each device, the output curves at fixed $|V_{GS}|\geq\SI{1.48}{\volt}$ and the transfer curves at fixed $|V_{DS}|\geq\SI{1.48}{\volt}$.}
  \label{fig:summary-results}
\end{figure}

Both searches reduce the in-sample mean device RRMS relative to the re-simulated released cards (Figure~\ref{fig:summary-results}), from 0.275 to 0.228 with seven parameters and to 0.204 with 43 parameters. The 43-parameter search has the lowest all-device, nMOS, and pMOS means, and a lower RRMS than the seven-parameter search on 16 of 18 devices. The ranking changes on the high-bias subset, where the seven-parameter cards have the lower mean (0.074 vs.\ 0.076). Appendix~\ref{app:devices} lists the per-device values.

The best cards already present in the simulator banks are close to the final scores. The best stored-bank means are 0.2297 (seven parameters) and 0.2068 (43 parameters), compared with final means of 0.2284 and 0.2038. In each parameter set, local finite-difference refinement supplies the selected card for 7 of the 18 devices, and the remaining 11 selected cards are stored NGSpice starts. Under this sampling range and optimization budget, the synthetic bank provides most of the improvement, and the surrogate adds a smaller refinement. Appendix~\ref{app:reproducibility} gives the bank and endpoint scores.

\begin{figure}[t]
  \centering
  \includegraphics[width=0.85\linewidth]{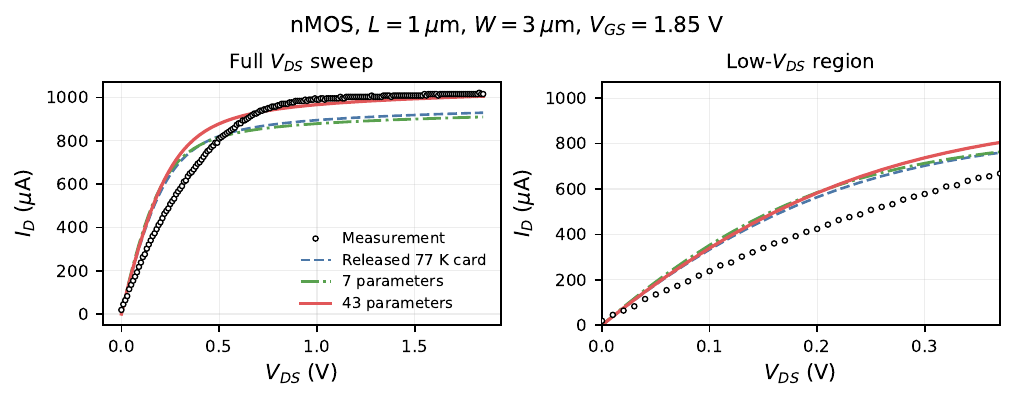}
  \caption{Output characteristic of the nMOS $L=\um{1}$, $W=\um{3}$ device at $V_{GS}=\SI{1.85}{\volt}$. Left: full $V_{DS}$ sweep. Right: $V_{DS}\leq\SI{0.37}{\volt}$. Lines are NGSpice re-simulations; markers are measurements.}
  \label{fig:example}
\end{figure}

Figure~\ref{fig:example} shows where the error occurs within one sweep. It plots the output characteristic of the nMOS $L=\um{1}$, $W=\um{3}$ device at $V_{GS}=\SI{1.85}{\volt}$. The 43-parameter card matches the measured saturation current, while the released and seven-parameter cards predict less current. All three cards predict more current than measured below $V_{DS}\approx\SI{0.5}{\volt}$, so the fitted cards improve the saturation region but leave the low-$V_{DS}$ mismatch in place.

\begin{figure}[t]
  \centering
  \includegraphics[width=0.85\linewidth]{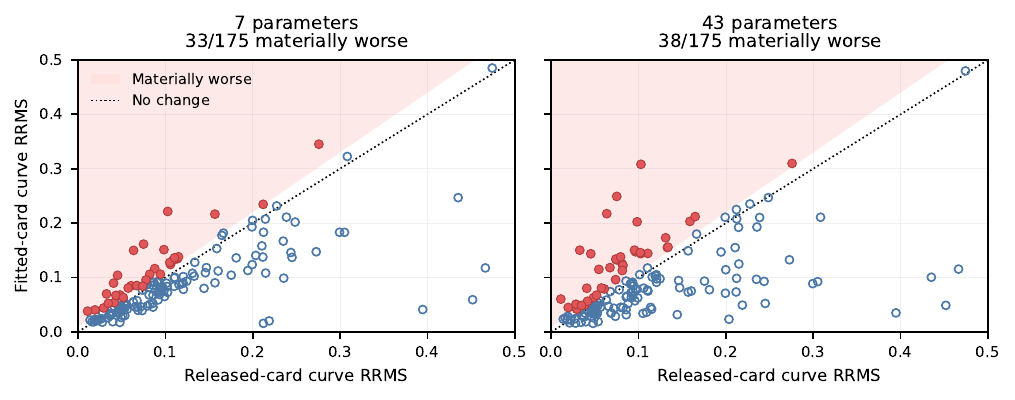}
  \caption{Per-curve RRMS, released card vs.\ fitted card. Each panel shows the 156 of 175 scored curves with released and fitted RRMS at most 0.5; the counts in the panel titles include all 175. Red markers and shading mark curves whose fitted RRMS exceeds the released-card RRMS by more than 0.01 and by more than 10\%. The flag uses a fixed threshold; no significance test is applied. One flagged seven-parameter curve lies outside the plotted range; all 38 flagged 43-parameter curves are shown.}
  \label{fig:curve-comparison}
\end{figure}

Figure~\ref{fig:curve-comparison} checks whether the aggregate improvement holds curve by curve. Most curves have lower RRMS with the fitted cards, but neither search improves every curve: 33 of 175 curves with seven parameters and 38 of 175 with 43 parameters have materially higher RRMS than with the released cards. The complete overlays in Appendix~\ref{app:curves} show these local changes for all 198 curves.

\section{Discussion}

The bank covers the bounded parameter space coarsely, and the surrogate gives local search directions between those samples. NGSpice confirmation keeps surrogate-predicted scores out of the reported results, but surrogate error can still change the optimization path and the shortlist of candidates. The 43-parameter surrogate has a larger approximation error than the seven-parameter surrogate (mean absolute surrogate--NGSpice difference in device RRMS of 0.0030 vs.\ 0.00035 at the selected cards), yet its confirmed cards have the lower extraction error. The surrogate provides local search evaluations, and the compact model in NGSpice is still used for every final evaluation.

Learning the forward map also avoids the need for a unique target parameter vector for each measured device. The synthetic bank directly supplies parameter vectors and their NGSpice responses for supervised training. During extraction, a candidate card is scored by how well its simulated currents match the measurement, and no assumed true parameter vector is used. The fitted parameters remain non-unique. The parameter values are therefore fit results and are not direct measurements of device physics.

The expanded parameter set gives more directions in which to match the measurements, but the improvement is uneven across bias: its overall mean is lower, while its high-bias mean is slightly higher than that of the seven-parameter fit. The objective averages residuals over complete curves, so improvement on many points can outweigh a larger error in a smaller operating region. The low-$V_{DS}$ mismatch in Figure~\ref{fig:example} is one example, and Figure~\ref{fig:curve-comparison} shows that curve-level regressions also occur on other devices. This experiment does not separate whether these residuals come from the parameter bounds, interactions among parameters, the objective weighting, or limits of the BSIM4 model form at \SI{77}{\kelvin}.

Output and transfer characteristics are fitted together, so each candidate is evaluated across the measured DC response. With equal curve weights, the score measures agreement averaged over all curves. A particular circuit operates over a narrower set of bias points, and a circuit-specific extraction could give more weight to those points. The difference between the overall and high-bias rankings shows that this weighting can change which card is preferred.

Surrogate-guided refinement reduces the mean RRMS by 0.0012 (seven parameters) and 0.0030 (43 parameters) beyond the best card in the simulator bank. The bank provides the broad search  and the surrogate allows continuous proposals between those samples. The experiment does not determine whether this is because the bank is already dense near low-error regions, because the local objective has shallow or competing directions, or because surrogate error limits the refinement.

\section{Limitations}
\label{sec:limitations}

The measured curves are used as fitting targets, and the same 18 geometries are used for sensitivity screening and surrogate training. No geometry is held out, so the results measure in-sample extraction for these 18 geometries; prediction for other channel dimensions or temperatures is not tested. The results also depend on one training seed, the chosen parameter bounds, the fixed curve-inclusion rule, and the weighting across curves and devices. Measurement uncertainty and repeatability are not propagated into the fit or the curve-regression flags. The study covers DC behavior for one process, temperature, and corner; dynamic response, noise, variability, reliability, and consistency between separately fitted geometries are not evaluated. The output is a set of fitted DC model cards for the 18 measured geometries.

\section{Conclusion}

A forward NGSpice surrogate supports bounded refinement of \SI{77}{\kelvin} BSIM4 cards while preserving standard NGSpice model cards. Across 18 measured SKY130 geometries, in-sample mean RRMS falls from 0.275 for re-simulated released cards to 0.228 (seven parameters) and 0.204 (43 parameters). The seven-parameter cards have the lower high-bias mean, and 33 (seven parameters) and 38 (43 parameters) of the 175 curves have materially higher RRMS than with the released cards. In this setting, the surrogate allows continuous local search without NGSpice in the optimization loop, but selection from the simulator bank accounts for most of the improvement. Final NGSpice confirmation and per-curve evaluation identified these regressions.

\begin{ack}
We thank Faith Beall for helpful discussions that clarified the Mystic extraction flow used in her earlier work~\cite{beall2026sky130}, and Ashish Rimal for his guidance on the SKY130 PDK. This material is based upon work supported by the U.S. Department of Energy (DOE), Office of Science, Offices of High Energy Physics and Basic Energy Sciences, as part of the Co-design and Heterogeneous Integration in Microelectronics for Extreme Environments (CHIME) Microelectronics Science Research Center (MSRC) (DOE Lab Announcement No. LAB 24-3320). Fermilab is managed by Fermi Forward Discovery Group, LLC, under Contract No. 89243024CSC000002. This work was also supported by the DOE Office of Science, Advanced Scientific Computing Research, as part of the Transformational Model Consortium (DOE Lab Announcement No. LAB 26-3560). O.S. was supported by a Graduate Instrumentation Research Award (GIRA) through the Coordinating Panel for Advanced Detectors (CPAD), funded by the DOE Office of High Energy Physics.
\end{ack}

\clearpage
\bibliographystyle{plainnat}
\bibliography{references}

@misc{beall2026sky130,
  author        = {Beall, F. and Rimal, A. and Seidel, O. and Mei, Y. and McDonald, A. D. and Parmaksiz, I. and Chirayath, V. A. and Asaadi, J. and Braga, D. and Battat, J. B. R.},
  title         = {{DC Cryogenic Modeling of Open-Source SkyWater 130 nm MOSFETs at 77 K Using BSIM4}},
  year          = {2026},
  eprint        = {2604.21625},
  archiveprefix = {arXiv},
  primaryclass  = {cond-mat.mes-hall},
  doi           = {10.48550/arXiv.2604.21625},
  url           = {https://arxiv.org/abs/2604.21625},
  note          = {arXiv:2604.21625v2}
}

@misc{cryopdkrepo,
  author  = {Gzamouranis, Olivia},
  title   = {{CryoPDK\_Skywater130nm\_ML: 77 K SKY130 Measurement, Simulation, and Model-Card Files}},
  year    = {2026},
  version = {39b1e518e25120104225b8fa19f4cfc61a6766b3},
  url     = {https://github.com/ogzamour/CryoPDK_Skywater130nm_ML/tree/39b1e518e25120104225b8fa19f4cfc61a6766b3}
}

@article{beckers2018cryogenic,
  author  = {Beckers, Arnout and Jazaeri, Farzan and Enz, Christian},
  title   = {{Cryogenic MOS Transistor Model}},
  journal = {IEEE Transactions on Electron Devices},
  year    = {2018},
  volume  = {65},
  number  = {9},
  pages   = {3617--3625},
  doi     = {10.1109/TED.2018.2854701},
  url     = {https://doi.org/10.1109/TED.2018.2854701}
}

@article{bohuslavskyi2019swing,
  author  = {Bohuslavskyi, H. and Jansen, A. G. M. and Barraud, S. and Barral, V. and Casse, M. and Le Guevel, L. and Jehl, X. and Hutin, L. and Bertrand, B. and Billiot, G. and Pillonnet, G. and Arnaud, F. and Galy, P. and De Franceschi, S. and Vinet, M. and Sanquer, M.},
  title   = {{Cryogenic Subthreshold Swing Saturation in FD-SOI MOSFETs Described With Band Broadening}},
  journal = {IEEE Electron Device Letters},
  year    = {2019},
  volume  = {40},
  number  = {5},
  pages   = {784--787},
  doi     = {10.1109/LED.2019.2903111},
  url     = {https://doi.org/10.1109/LED.2019.2903111}
}

@article{beckers2020swing,
  author  = {Beckers, Arnout and Jazaeri, Farzan and Enz, Christian},
  title   = {{Theoretical Limit of Low Temperature Subthreshold Swing in Field-Effect Transistors}},
  journal = {IEEE Electron Device Letters},
  year    = {2020},
  volume  = {41},
  number  = {2},
  pages   = {276--279},
  doi     = {10.1109/LED.2019.2963379},
  url     = {https://doi.org/10.1109/LED.2019.2963379}
}

@article{incandela2018characterization,
  author  = {Incandela, Rosario M. and Song, Lin and Homulle, Harald and Charbon, Edoardo and Vladimirescu, Andrei and Sebastiano, Fabio},
  title   = {{Characterization and Compact Modeling of Nanometer CMOS Transistors at Deep-Cryogenic Temperatures}},
  journal = {IEEE Journal of the Electron Devices Society},
  year    = {2018},
  volume  = {6},
  pages   = {996--1006},
  doi     = {10.1109/JEDS.2018.2821763},
  url     = {https://doi.org/10.1109/JEDS.2018.2821763}
}

@article{gatti2024small,
  author  = {Gatti, Alberto and Tavernier, Filip},
  title   = {{Cryogenic Small Dimension Effects and Design-Oriented Scalable Compact Modeling of a 65-nm CMOS Technology}},
  journal = {IEEE Journal of the Electron Devices Society},
  year    = {2024},
  volume  = {12},
  pages   = {369--378},
  doi     = {10.1109/JEDS.2024.3394167},
  url     = {https://doi.org/10.1109/JEDS.2024.3394167}
}

@article{beckers2019fdsoi,
  author  = {Beckers, Arnout and Jazaeri, Farzan and Bohuslavskyi, Heorhii and Hutin, Louis and De Franceschi, Silvano and Enz, Christian},
  title   = {{Characterization and Modeling of 28-nm FDSOI CMOS Technology Down to Cryogenic Temperatures}},
  journal = {Solid-State Electronics},
  year    = {2019},
  volume  = {159},
  pages   = {106--115},
  doi     = {10.1016/j.sse.2019.03.033},
  url     = {https://doi.org/10.1016/j.sse.2019.03.033}
}

@article{braga2021hep,
  author  = {Braga, Davide and Li, Shaorui and Fahim, Farah},
  title   = {{Cryogenic Electronics Development for High-Energy Physics: An Overview of Design Considerations, Benefits, and Unique Challenges}},
  journal = {IEEE Solid-State Circuits Magazine},
  year    = {2021},
  volume  = {13},
  number  = {2},
  pages   = {36--45},
  doi     = {10.1109/MSSC.2021.3072804},
  url     = {https://doi.org/10.1109/MSSC.2021.3072804}
}

@article{potocnik2025control,
  author  = {Poto{\v{c}}nik, Anton},
  title   = {{How to Scale the Electronic Control Systems of a Quantum Computer}},
  journal = {Nature Electronics},
  year    = {2025},
  volume  = {8},
  number  = {1},
  pages   = {3--4},
  doi     = {10.1038/s41928-024-01331-9},
  url     = {https://doi.org/10.1038/s41928-024-01331-9}
}

@article{kao2022bsimcmg,
  author  = {Kao, Ming-Yen and Chavez, Fredo and Khandelwal, Sourabh and Hu, Chenming},
  title   = {{Deep Learning-Based BSIM-CMG Parameter Extraction for 10-nm FinFET}},
  journal = {IEEE Transactions on Electron Devices},
  year    = {2022},
  volume  = {69},
  number  = {8},
  pages   = {4765--4768},
  doi     = {10.1109/TED.2022.3181536},
  url     = {https://doi.org/10.1109/TED.2022.3181536}
}

@article{chavez2023global,
  author  = {Chavez, Fredo and Tung, Chien-Ting and Kao, Ming-Yen and Hu, Chenming and Chen, Jen-Hao and Khandelwal, Sourabh},
  title   = {{Deep Learning-Based I--V Global Parameter Extraction for BSIM-CMG}},
  journal = {Solid-State Electronics},
  year    = {2023},
  volume  = {209},
  pages   = {108766},
  doi     = {10.1016/j.sse.2023.108766},
  url     = {https://doi.org/10.1016/j.sse.2023.108766}
}

@article{cranmer2020frontier,
  author  = {Cranmer, Kyle and Brehmer, Johann and Louppe, Gilles},
  title   = {{The Frontier of Simulation-Based Inference}},
  journal = {Proceedings of the National Academy of Sciences},
  year    = {2020},
  volume  = {117},
  number  = {48},
  pages   = {30055--30062},
  doi     = {10.1073/pnas.1912789117},
  url     = {https://doi.org/10.1073/pnas.1912789117}
}

@article{hesthaven2018nonintrusive,
  author  = {Hesthaven, Jan S. and Ubbiali, Stefano},
  title   = {{Non-Intrusive Reduced Order Modeling of Nonlinear Problems Using Neural Networks}},
  journal = {Journal of Computational Physics},
  year    = {2018},
  volume  = {363},
  pages   = {55--78},
  doi     = {10.1016/j.jcp.2018.02.037},
  url     = {https://doi.org/10.1016/j.jcp.2018.02.037}
}

@article{virtanen2020scipy,
  author  = {Virtanen, Pauli and Gommers, Ralf and Oliphant, Travis E. and Haberland, Matt and Reddy, Tyler and Cournapeau, David and Burovski, Evgeni and Peterson, Pearu and Weckesser, Warren and Bright, Jonathan and van der Walt, St{\'e}fan J. and Brett, Matthew and Wilson, Joshua and Millman, K. Jarrod and Mayorov, Nikolay and Nelson, Andrew R. J. and Jones, Eric and Kern, Robert and Larson, Eric and Carey, C. J. and Polat, {\.{I}}lhan and Feng, Yu and Moore, Eric W. and VanderPlas, Jake and Laxalde, Denis and Perktold, Josef and Cimrman, Robert and Henriksen, Ian and Quintero, E. A. and Harris, Charles R. and Archibald, Anne M. and Ribeiro, Ant{\^o}nio H. and Pedregosa, Fabian and van Mulbregt, Paul and SciPy 1.0 Contributors},
  title   = {{SciPy 1.0: Fundamental Algorithms for Scientific Computing in Python}},
  journal = {Nature Methods},
  year    = {2020},
  volume  = {17},
  number  = {3},
  pages   = {261--272},
  doi     = {10.1038/s41592-019-0686-2},
  url     = {https://doi.org/10.1038/s41592-019-0686-2}
}

@manual{bsim4manual,
  author       = {{BSIM Group}},
  title        = {{BSIM4 4.8.3 MOSFET Model: User's Manual}},
  organization = {University of California, Berkeley},
  year         = {2025},
  url          = {https://bsim.berkeley.edu/models/bsim4/}
}

@misc{ngspice41,
  author  = {{The ngspice team}},
  title   = {{ngspice, Version 41}},
  year    = {2023},
  url     = {https://sourceforge.net/projects/ngspice/files/ng-spice-rework/old-releases/41/}
}


\clearpage
\appendix

\section{Reproducibility details}
\label{app:reproducibility}

\subsection{Fixed configuration}

Table~\ref{tab:configuration} lists the settings shared by the seven- and 43-parameter experiments. Both use seed 20260818. Training and extraction ran with PyTorch 2.12 on the MPS backend of an Apple M2 Max with 64 GB of unified memory. All final current vectors were generated by conda-forge NGSpice 41 on the CPU. Surrogate training took 1,664.2 s for seven parameters and 1,691.8 s for 43 parameters; extraction with the fixed surrogate took 31.5 s and 235.8 s, respectively. Generating the 43-parameter NGSpice Latin hypercube took 2,145.1 s. The seven-parameter run reused an existing simulator bank, for which no separate generation time was recorded. All reported values come from one seeded training and extraction run per parameter set; no across-seed error bars are reported.

For both parameter sets, each device sends four surrogate endpoints, their four stored NGSpice starts, and the anchor to NGSpice, giving nine confirmations after de-duplication. Seven of the 18 selected cards are endpoints and 11 are stored starts. The best stored-bank means are 0.22967 (seven parameters) and 0.20678 (43 parameters), compared with final means of 0.22844 and 0.20380. At the selected cards, the mean absolute surrogate--NGSpice difference in device RRMS is 0.00035 (maximum 0.00158) for seven parameters and 0.00299 (maximum 0.01410) for 43 parameters.

The reported baseline always uses the unmodified released cards. In the 43-parameter layout, writing every parameter at its read-back anchor value gives a mean RRMS of 0.27792. The unmodified cards give 0.27505, with a maximum device-level difference of 0.01111. The anchor is included only as a confirmation candidate and is not selected for any reported result.

\begin{table}[t]
  \caption{Surrogate and extraction configuration.}
  \label{tab:configuration}
  \centering
  \small
  \begin{tabular}{@{}ll@{}}
    \toprule
    Item & Value \\
    \midrule
    Devices & 8 nMOS, 10 pMOS; 18 measured geometries \\
    Curves per device & 5 output, 6 transfer; 2,046 points \\
    Synthetic data & 10,000 LHC rows per geometry and parameter set \\
    Surrogate inputs & parameters, polarity, scaled $\log L$, $\log W$ \\
    Hidden layers & $4\times1024$; GELU \\
    Split / batch size & 80\%/10\%/10\% per device; 512 rows \\
    Optimizer & AdamW; initial LR $10^{-3}$; weight decay $10^{-5}$ \\
    Schedule / epochs & cosine decay; at most 200; patience 25 \\
    Hybrid loss & signed-log MSE + linear Huber ($\beta=1$), weights 1:1 \\
    Geometry scaling & polarity and linearly scaled $\log L$, $\log W$ in $[-1,1]$ \\
    Extraction starts & 8 diverse NGSpice LHC vectors \\
    Start separation & RMS active-coordinate distance $\geq0.35$ \\
    Least squares & TRF, two-point Jacobian, step 0.02 \\
    Search limits & 100 function evaluations; bounds $[-1,1]$ \\
    Prior & $10^{-4}$ penalty toward released parameters \\
    NGSpice confirmation & 4 endpoints, 4 starts, anchor \\
    \bottomrule
  \end{tabular}
\end{table}

Let $s(I)=\operatorname{sign}(I)\log(1+|I|/10^{-9}\,\mathrm{A})$ and $z_j=[s(I_j)-\mu_j]/\sigma_j$, where $\mu_j$ and $\sigma_j$ are computed pointwise over the synthetic bank. For device $d$ and point $j$, the linear-current scale $a_{d,j}$ is the 99th percentile of $|I_j|$ within that curve, bounded below by both $10^{-15}$~A and $10^{-6}$ times the largest curve scale of that device. With uniform point weights, the forward loss is
\begin{equation}
  \mathcal L_{\mathrm{fwd}}=
  \operatorname{mean}_j(\hat z_j-z_j)^2+
  \operatorname{mean}_j H_1\!\left(
  \frac{\hat I_j-I_j}{a_{d,j}}\right),
  \label{eq:forward-loss}
\end{equation}
where $H_1$ is the smooth-$L_1$ loss with transition 1. The checkpoint with the best validation loss is kept, with patience 25 and a 200-epoch limit; both runs reach the limit. On the 18,000-example synthetic test split, the total loss is $9.14\times10^{-7}$ for seven parameters and $1.06\times10^{-3}$ for 43 parameters. The signed-log MSE components are $6.37\times10^{-7}$ and $9.96\times10^{-4}$, and the linear Huber components are $2.77\times10^{-7}$ and $6.70\times10^{-5}$. Because the normalization statistics use the full synthetic bank, these values measure the interpolation error of the surrogate on simulated data.

For active extraction coordinates $x_A\in[-1,1]^{D_A}$, where $A$ is the device-specific active set and $D_A$ its size, the least-squares residual gives every admitted curve equal weight. The data term is
\begin{equation}
  \mathcal L_{\mathrm{inv}}(x_A)=\frac{1}{K}\sum_{k=1}^{K}
  \frac{1}{N_k}\sum_{i=1}^{N_k}
  \left(\frac{\hat I_{k,i}(x_A)-I_{k,i}}
  {N_k^{-1}\sum_j|I_{k,j}|}\right)^2
  +\frac{10^{-4}}{D_A}\lVert x_A\rVert_2^2.
  \label{eq:inverse-loss}
\end{equation}

\subsection{Curve handling and scoring}

The curve order is output sweeps at fixed $|V_{GS}|=0.37$, 0.74, 1.11, 1.48, \SI{1.85}{\volt}, followed by transfer sweeps at fixed $|V_{DS}|=0.01$, 0.37, 0.74, 1.11, 1.48, \SI{1.85}{\volt}. The implementation keeps the released pMOS negative sweep direction, device multiplicity, parasitic instance values, and model-bin selection. Measured currents before the last exact zero are set to zero. Leading pMOS points are trimmed at the released device-specific current floors. The exact $(I_{VG},I_{VD})$ floors in amperes, in the order of Table~\ref{tab:device-results}, are
\begin{quote}\small
$(10^{-10},0)$, $(6\times10^{-10},0)$, $(6\times10^{-6},10^{-7})$,
$(10^{-7},5\times10^{-8})$, $(0,0)$, $(6\times10^{-10},0)$,
$(1.2\times10^{-9},0.75)$, $(1.8\times10^{-9},0)$,
$(10^{-9},0)$, and $(4.5\times10^{-9},0)$.
\end{quote}
For nMOS, a curve is admitted when the trimmed mean $|I_{\mathrm{meas}}|$ exceeds $3\times10^{-7}$~A for output curves or $2\times10^{-7}$~A for transfer curves; the $L=W=\um{100}$ device uses $3\times10^{-5}$~A and $9\times10^{-10}$~A. For pMOS, the final simulated magnitude of the released card must exceed $10^{-10}$~A for output curves or $10^{-11}$~A for transfer curves. We freeze this released-card inclusion set before fitting. The 23 excluded curves are the $|V_{GS}|=\SI{0.37}{\volt}$ output curves of all eight nMOS devices; the $|V_{GS}|=\SI{0.37}{\volt}$ output curves of every pMOS device except $L=\um{0.35}$, $W=\um{0.55}$; and the $|V_{GS}|=\SI{0.74}{\volt}$ output curves of the pMOS devices with $(L,W)=(0.5,0.42)$, $(2,5)$, $(4,7)$, $(8,0.84)$, $(8,1.6)$, and $(8,5)$~\si{\micro\metre}. No transfer curve is excluded. Equation~\ref{eq:rrms} is evaluated on the same cleaned and trimmed spans for every method.

\subsection{Parameter sweep details}

To sweep the seven selected parameters, we first added a multiplier to each of them in the room-temperature PDK. Each multiplier is defined as 1 in the global section of the PDK and can be set externally for each sweep and each device. We then chose a range for each nominal multiplier from the expected behavior of the corresponding parameter. The Latin-hypercube script checks each range by confirming that NGSpice produces a valid I--V curve at both ends of the range. The validated ranges are passed to the Latin-hypercube Monte Carlo framework, which generates the sampled multiplier values and sends them to NGSpice to produce the training data. The two previously extracted PDKs are used only as the baseline; their extracted multiplier values were not used to set the ranges, so this procedure does not depend on a prior extraction and can be applied to a PDK that has not been optimized before.

\section{Parameter sets}
\label{app:parameters}

The original seven-parameter set is
\begin{center}
  \param{VTH0}, \param{U0}, \param{NFACTOR}, \param{VSAT},
  \param{DELTA}, \param{RDSW}, \param{ETA0}.
\end{center}
For the expanded set, 115 candidate continuous DC-current parameters were read from the nMOS and pMOS BSIM4 cards. Each candidate was perturbed by a nominal 10\% span and simulated on all 11 curves. For a parameter whose released value is zero, the span uses the median nonzero magnitude within that polarity, with the median across both polarities as the fallback. For curve $k$, the normalized response to perturbation $p$ is
\begin{equation}
  S_{p,k}=
  \frac{\operatorname{RMS}(I_{p,k}-I_{\mathrm{rel},k})}
  {\max\!\left(\operatorname{mean}|I_{\mathrm{meas},k}|,
  \max(10^{-15}\,\mathrm{A},10^{-6}I_{\mathrm{peak}})\right)},
  \label{eq:sensitivity}
\end{equation}
where $I_{\mathrm{peak}}$ is the largest measured current magnitude for that device. A parameter is kept if $S_{p,k}\geq10^{-4}$ for either perturbation direction on any curve and geometry. The original seven parameters are always kept as a nested subset. Non-finite perturbations are retried at 5\%, 2.5\%, and 1\% of the nominal span; the released value and the largest finite perturbation in each available direction define a device-specific, possibly asymmetric interval. Parameters with categorical behavior, geometry-selection roles, capacitance- or noise-only effects, unsafe discontinuities, exact aliases, or no response above the threshold are not kept. Exact per-device bounds are listed in \param{parameter\_sensitivity\_summary.json}. The union over all 18 devices contains the following 43 parameters:
\begin{quote}\small
\param{TOXE}, \param{TOXM}, \param{XJ}, \param{NDEP}, \param{NSD},
\param{DWG}, \param{RSH}, \param{VTH0}, \param{K1}, \param{K2},
\param{K3}, \param{DVT0}, \param{DVT1}, \param{DVT1W}, \param{W0},
\param{K3B}, \param{VFB}, \param{LPE0}, \param{LPEB}, \param{VSAT},
\param{UA}, \param{UB}, \param{RDSW}, \param{PRWG}, \param{WR},
\param{U0}, \param{A0}, \param{A2}, \param{AGS}, \param{B0},
\param{B1}, \param{VOFF}, \param{NFACTOR}, \param{CIT}, \param{CDSCD},
\param{ETA0}, \param{DSUB}, \param{PCLM}, \param{PDIBLC2},
\param{DROUT}, \param{DELTA}, \param{ALPHA0}, and \param{BETA0}.
\end{quote}
A parameter in the all-device union is an input for both polarities. Parameters that are absent from a card, fixed at zero without a safe nonzero scale, or without a non-degenerate safe interval are held at their released value for that device. This leaves 38 active dimensions for nMOS and 39 or 40 for pMOS within a common 43-feature input layout.

\section{Per-device RRMS}
\label{app:devices}

\begin{table}[h]
  \caption{Per-device RRMS from NGSpice re-simulations. ``Released'' is the released \SI{77}{\kelvin} parameter vector re-simulated with the same simulator and scorer.}
  \label{tab:device-results}
  \centering
  \small
  \sisetup{round-mode=none}
  \begin{tabular}{@{}l
                  S[table-format=3.2]
                  S[table-format=3.2]
                  S[table-format=1.4]
                  S[table-format=1.4]
                  S[table-format=1.4]@{}}
    \toprule
    Type & {$L$ (\si{\micro\metre})} & {$W$ (\si{\micro\metre})} & {Released} & {7 param.} & {43 param.} \\
    \midrule
    nMOS & 0.15 & 1.6  & 0.0595 & 0.0351 & 0.0380 \\
    nMOS & 0.19 & 7    & 0.0967 & 0.0436 & 0.0346 \\
    nMOS & 0.25 & 1.6  & 0.0983 & 0.0487 & 0.0438 \\
    nMOS & 1    & 1.6  & 0.1283 & 0.0637 & 0.0648 \\
    nMOS & 1    & 3    & 0.1599 & 0.1102 & 0.0657 \\
    nMOS & 8    & 1.6  & 0.1474 & 0.0809 & 0.0766 \\
    nMOS & 20   & 0.64 & 0.1283 & 0.1161 & 0.1083 \\
    nMOS & 100  & 100  & 0.1403 & 0.1324 & 0.1317 \\
    \midrule
    pMOS & 0.35 & 0.55 & 0.7270 & 0.6335 & 0.6203 \\
    pMOS & 0.35 & 1.6  & 0.3776 & 0.3348 & 0.2691 \\
    pMOS & 0.35 & 5    & 0.3291 & 0.2882 & 0.2813 \\
    pMOS & 0.5  & 0.42 & 0.4560 & 0.3670 & 0.3161 \\
    pMOS & 0.5  & 0.64 & 0.3498 & 0.3282 & 0.3108 \\
    pMOS & 2    & 5    & 0.1907 & 0.1783 & 0.1670 \\
    pMOS & 4    & 7    & 0.2702 & 0.2271 & 0.1855 \\
    pMOS & 8    & 0.84 & 0.4340 & 0.3813 & 0.3343 \\
    pMOS & 8    & 1.6  & 0.5433 & 0.4768 & 0.3940 \\
    pMOS & 8    & 5    & 0.3146 & 0.2662 & 0.2263 \\
    \midrule
    \multicolumn{3}{@{}l}{All-device mean} & 0.2751 & 0.2284 & 0.2038 \\
    \bottomrule
  \end{tabular}
\end{table}
\FloatBarrier

\section{Alternate modeling method}
\label{app:altMeth}

The alternate method has two stages: a forward surrogate and a self-supervised inverse-extraction network. In the first stage, the forward surrogate is trained by supervised regression to map seven SPICE model parameters to a 370-dimensional I--V response. The surrogate has three or four fully connected hidden layers with LeakyReLU activations, $L_2$ weight regularization, and optional dropout. Layer width, dropout rate, and learning rate are selected with Keras Tuner, and training uses the Adam optimizer with an exponentially decaying learning rate.

In the second stage, an inverse network maps a measured I--V response to the seven model parameters. Parameter labels are not used to train this stage. Instead, the predicted parameters pass through the frozen forward surrogate, and the reconstructed I--V curve is compared with the input curve using the mean squared error in log-current space. Gradients through the frozen surrogate update the inverse network while the forward model stays fixed. The forward and inverse networks use the same general dense-network design but are tuned separately.

This result covers only the nMOS device with $L=\um{0.15}$ and $W=\um{1.6}$ at fixed $|V_{GS}|=\SI{1.48}{\volt}$. Its training set contains approximately 56,000 I--V curves generated with the same Latin-hypercube sampling procedure. It is a single-device comparison and is not included in the 18-device aggregate results.

\begin{figure}[h]
    \centering
    \includegraphics[width=0.75\linewidth]{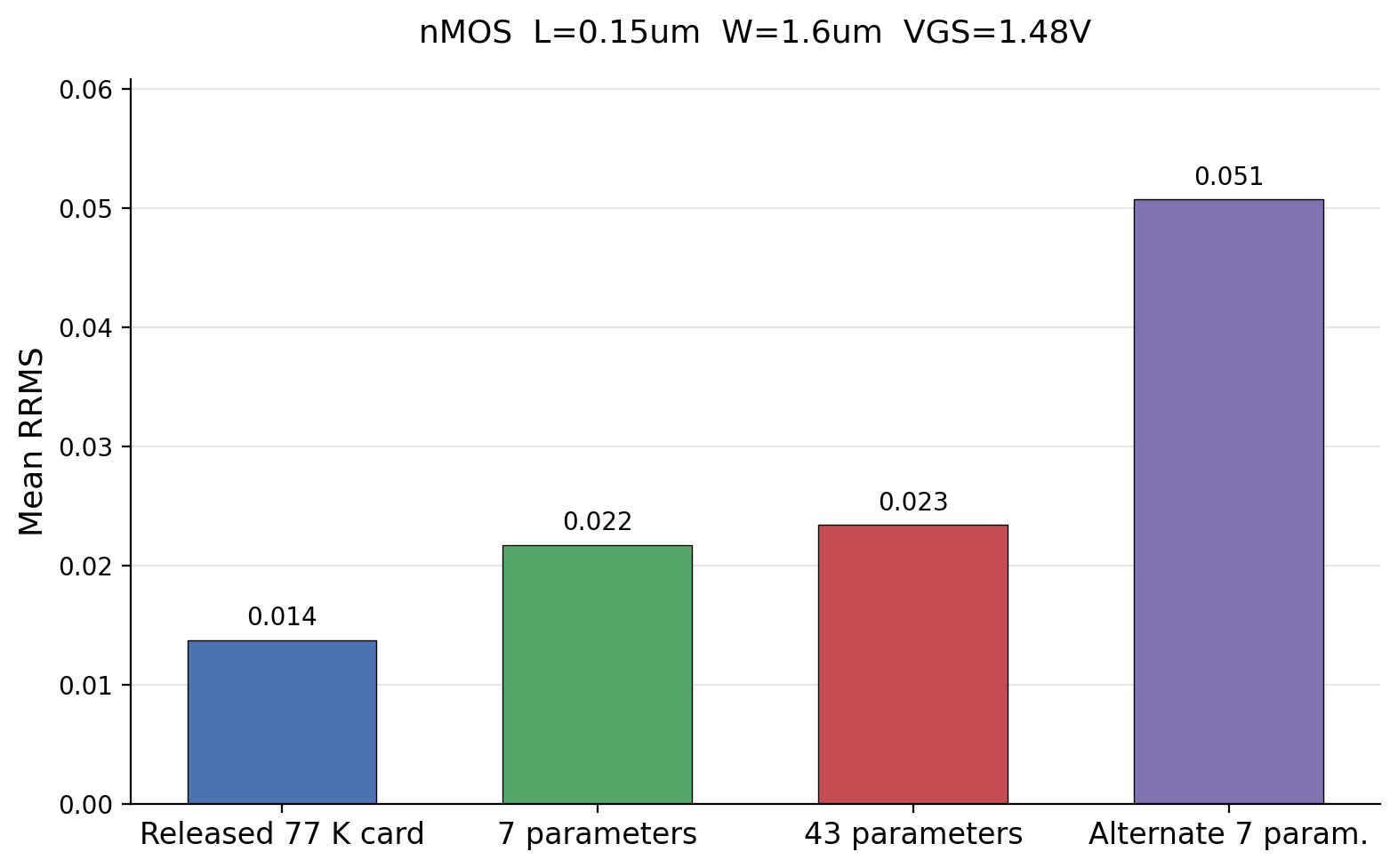}
    \caption{RRMS for the nMOS $L=\um{0.15}$, $W=\um{1.6}$ device at fixed $|V_{GS}|=\SI{1.48}{\volt}$ for the main and alternate methods.}
    \label{fig:altModelComp}
\end{figure}

\FloatBarrier

\section{Complete I--V overlays}
\label{app:curves}

The following figures show all 198 measured curves: 11 sweeps for each of the 18 geometries. The frozen scorer admits 175 of these curves, as described in Appendix~\ref{app:reproducibility}; the excluded curves are also plotted. Markers are measured data. Dashed, dash-dotted, and solid lines are NGSpice re-simulations of the released, seven-parameter, and 43-parameter cards. Color denotes fixed $|V_{GS}|$ in the output panel and fixed $|V_{DS}|$ in the transfer panels. Current magnitudes are plotted for both polarities. The linear panels show the strong-inversion shape, and the logarithmic transfer panel shows the subthreshold and near-off regions.

\newcommand{\deviceiv}[1]{%
  \includegraphics[width=\linewidth]{iv/#1.pdf}\par\vspace{1.0ex}}

\begin{figure}[p]
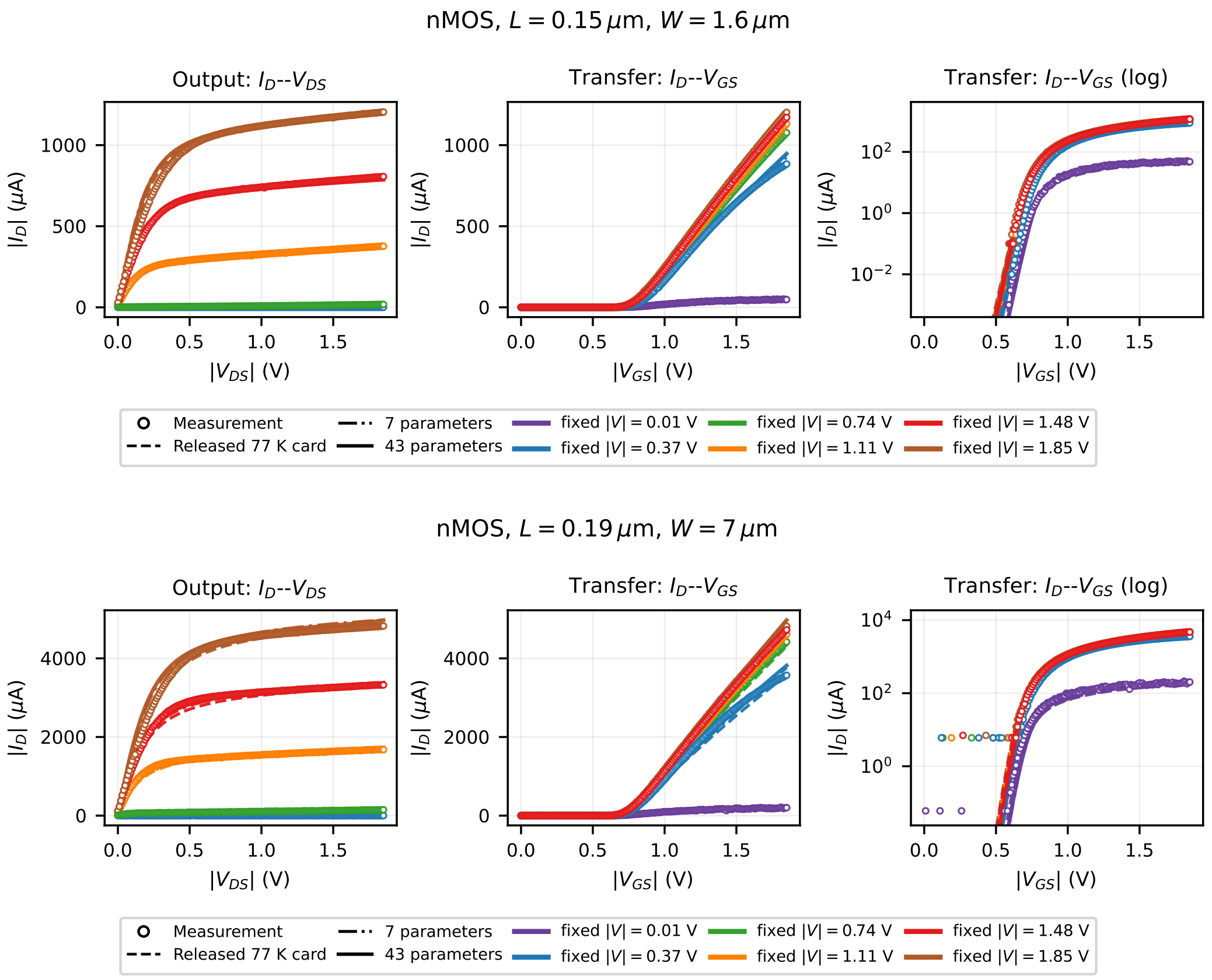
\centering
  \deviceiv{nmos_L0p15_W1p6}
  \deviceiv{nmos_L0p19_W7}
  \caption{Complete I--V overlays for nMOS devices (1 of 4).}
\end{figure}
\clearpage

\begin{figure}[p]\centering
  \deviceiv{nmos_L0p25_W1p6}
  \deviceiv{nmos_L1_W1p6}
  \caption{Complete I--V overlays for nMOS devices (2 of 4).}
\end{figure}
\clearpage

\begin{figure}[p]\centering
  \deviceiv{nmos_L1_W3}
  \deviceiv{nmos_L8_W1p6}
  \caption{Complete I--V overlays for nMOS devices (3 of 4).}
\end{figure}
\clearpage

\begin{figure}[p]\centering
  \deviceiv{nmos_L20_W0p64}
  \deviceiv{nmos_L100_W100}
  \caption{Complete I--V overlays for nMOS devices (4 of 4).}
\end{figure}
\clearpage

\begin{figure}[p]\centering
  \deviceiv{pmos_L0p35_W0p55}
  \deviceiv{pmos_L0p35_W1p6}
  \caption{Complete I--V overlays for pMOS devices (1 of 5).}
\end{figure}
\clearpage

\begin{figure}[p]\centering
  \deviceiv{pmos_L0p35_W5}
  \deviceiv{pmos_L0p5_W0p42}
  \caption{Complete I--V overlays for pMOS devices (2 of 5).}
\end{figure}
\clearpage

\begin{figure}[p]\centering
  \deviceiv{pmos_L0p5_W0p64}
  \deviceiv{pmos_L2_W5}
  \caption{Complete I--V overlays for pMOS devices (3 of 5).}
\end{figure}
\clearpage

\begin{figure}[p]\centering
  \deviceiv{pmos_L4_W7}
  \deviceiv{pmos_L8_W0p84}
  \caption{Complete I--V overlays for pMOS devices (4 of 5).}
\end{figure}
\clearpage

\begin{figure}[p]\centering
  \deviceiv{pmos_L8_W1p6}
  \deviceiv{pmos_L8_W5}
  \caption{Complete I--V overlays for pMOS devices (5 of 5).}
\end{figure}

\end{document}